%% file: main.tex
\documentclass[sigconf, screen]{acmart} 
\acmConference[ICSE 2024]{46th International Conference on Software Engineering}{April 2024}{Lisbon, Portugal}
\usepackage[normalem]{ulem}
\usepackage[utf8]{inputenc}
\usepackage{amsmath}  
\usepackage{graphicx}
\usepackage{comment}
\usepackage{multirow}
\usepackage{enumitem}
\usepackage{array}
\usepackage{pifont}
\usepackage{xcolor}
\usepackage{balance}
\usepackage{mathtools}
\usepackage{xspace}
\usepackage{url}
\usepackage{eucal}
\usepackage{amsfonts}
\usepackage{color}
\usepackage{booktabs}
\usepackage{bbding}
\usepackage{float}

\usepackage{graphicx}
\usepackage{hyperref}
\usepackage{url}
\usepackage{multirow}
\usepackage{comment}
\usepackage{colortbl}
\usepackage{graphicx}
\usepackage{subcaption}
\usepackage[linewidth=0.1pt]{mdframed}
\usepackage[linesnumbered,ruled,lined]{algorithm2e}

\usepackage{wrapfig,lipsum,booktabs}
\usepackage{subfloat}
\usepackage[export]{adjustbox}

\AtBeginDocument{%
  \providecommand\BibTeX{{%
    \normalfont B\kern-0.5em{\scshape i\kern-0.25em b}\kern-0.8em\TeX}}}

\newcommand{\revised}[1]{\textcolor{black}{#1}}
\newcommand{\revisedzx}[1]{\textcolor{black}{#1}}
\newcommand{\revisedtwo}[1]{\textcolor{black}{#1}}

\pdfpagewidth=\paperwidth
\pdfpageheight=\paperheight

\makeatletter
\newcommand{\showfontsize}{\f@size{} pt}
\newcommand\usemm[1]{%
  \strip@pt\dimexpr0.3514598\dimexpr #1\relax\relax mm%
}
\newcommand\usein[1]{%
  \strip@pt\dimexpr0.013837\dimexpr #1\relax\relax in%
}
\makeatother

\usepackage{graphicx}
\usepackage{hyperref}
\usepackage{url}
\usepackage{multirow}
\usepackage{comment}
\usepackage{colortbl}
\usepackage{graphicx}
\usepackage{tcolorbox}
\usepackage{arydshln}
\usepackage{algpseudocode}  
\usepackage{amsmath}  
\usepackage{caption} 
\usepackage{algpseudocode}
\usepackage{float}
\usepackage{soul}

\usepackage{hyperref}
\usepackage[hyphenbreaks]{breakurl}

\usepackage{xcolor}
\usepackage{enumitem}
\usepackage{amsmath}

\definecolor{dark-red}{RGB}{255,0,0}
\definecolor{dark-green}{RGB}{0,200,0}

\newboolean{showcomments}
\setboolean{showcomments}{true}
\ifthenelse{\boolean{showcomments}}

\acmDOI{}
\copyrightyear{2024} 
\acmYear{2024} 
\setcopyright{rightsretained} 
\acmConference[ICSE-NIER'24]{New Ideas and Emerging Results }{April 14--20, 2024}{Lisbon, Portugal}
\acmBooktitle{New Ideas and Emerging Results (ICSE-NIER'24), April 14--20, 2024, Lisbon, Portugal}\acmDOI{10.1145/3639476.3639762}
\acmISBN{979-8-4007-0500-7/24/04}
\begin{document}


\pagestyle{fancy}
\fancypagestyle{preContent}{
    \renewcommand\headrulewidth{0pt}
    \fancyfoot[C]{\thepage}
}
\pagestyle{preContent}
\pagenumbering{arabic}


\settopmatter{printacmref=false}

 \title{\revised{Large Language Model for Vulnerability Detection: \newline Emerging Results and Future Directions}}
 
\author{Xin Zhou, Ting Zhang, and David Lo}
\affiliation{
    \institution{School of Computing and Information Systems, Singapore Management University, Singapore}
    \country{}
}
\email{{xinzhou.2020, tingzhang.2019}@phdcs.smu.edu.sg, davidlo@smu.edu.sg}

\input{Sections/00_abstract}

\maketitle

\input{Sections/01_introduction}

\input{Sections/02_model}

\input{Sections/03_experiments}

\input{Sections/04_threats}
\input{Sections/related_work}

\input{Sections/05_future_work}

\input{Sections/06_conclusions}

\balance
\bibliographystyle{unsrt}
\bibliography{references}

\end{document}

%% file: Sections/00_abstract.tex
\begin{abstract}\label{abstract}

Previous learning-based vulnerability detection methods relied on either medium-sized pre-trained models or smaller neural networks from scratch.
Recent advancements in Large Pre-Trained Language Models (LLMs) have showcased remarkable few-shot learning capabilities in various tasks. 
However, the effectiveness of LLMs in detecting software vulnerabilities is largely unexplored. 
This paper aims to bridge this gap by exploring how LLMs perform with various prompts, particularly focusing on two state-of-the-art LLMs: GPT-3.5 and GPT-4. 
Our experimental results showed that GPT-3.5 achieves competitive performance with the prior state-of-the-art vulnerability detection approach and GPT-4 consistently outperformed the state-of-the-art.

\end{abstract}

%% file: Sections/01_introduction.tex
\vspace{-0.2cm}
\section{Introduction}

Software vulnerabilities are the prevalent issues in software systems, posing various risks such as the compromise of sensitive information~\cite{MSFlaw} and system failures~\cite{turner2008symantec}. 
To address this challenge, researchers have proposed machine learning (ML) and deep learning (DL) approaches for identifying vulnerabilities in source code~\cite{hanif2022vulberta,fu2022linevul,nguyen2022regvd,zhou2019devign}. 
While previous ML/DL-based vulnerability detection approaches have demonstrated promising results, they have primarily relied on either medium-size pre-trained models such as CodeBERT~\cite{CodeBERT,fu2022linevul} or training smaller neural networks (such as Graph Neural Networks~\cite{nguyen2022regvd}) from scratch.

Recent developments in Large Pre-Trained Language Models (LLMs) have demonstrated impressive few-shot learning capabilities across diverse tasks~\cite{xia2023keep,zhang2023cupid,zhang2023revisiting,weyssow2023exploring,zhou2023patchzero}.
However, the performance of LLMs on security-oriented tasks, particularly vulnerability detection, remains largely unexplored. 
Moreover, LLMs are gradually starting to be used in software engineering (SE), as seen in automated program repair~\cite{xia2023keep}. 
However, these studies predominantly focus on using LLMs for generation-based tasks. 
It remains unclear whether LLMs can be effectively utilized in classification tasks and outperform the medium-size pre-trained models such as CodeBERT, specifically in the vulnerability detection task.

To fill in the research gaps, this paper investigates the effectiveness of LLMs in identifying vulnerable codes, i.e., a critical classification task within the domain of security.
Furthermore, the efficacy of LLMs heavily relies on the quality of prompts (task descriptions and other relevant information) provided to the model. 
Thus, we explore and design diverse prompts to effectively apply LLMs for vulnerability detection.
We specifically studied two prominent state-of-the-art Large Language Models (LLMs): \textbf{GPT-3.5} and \textbf{GPT-4}, both serving as the foundational models for ChatGPT.
Our experimental results revealed that with appropriate prompts, GPT-3.5 achieves competitive performance with CodeBERT, and GPT-4 outperformed CodeBERT by 34.8\% in terms of Accuracy.
\revised{In summary, our contributions are as follows:}
\begin{itemize}[leftmargin=*]
\vspace{-0.1cm}

\item [$\bullet$] \revisedtwo{We conduct experiments with diverse prompts for LLMs, encompassing task and role descriptions, project information, and examples from Common Weakness Enumeration (CWE) and the training set. We recognize LLMs as promising models for vulnerability detection.}

\item [$\bullet$] \revisedtwo{ We pinpoint several promising future directions for leveraging LLMs in vulnerability detection, and we encourage the community to delve into these possibilities.}

\end{itemize}

%% file: Sections/02_model.tex
\vspace{-0.2cm}
\section{Proposed Approach}

\noindent \textbf{ChatGPT and In-Context Learning.}
ChatGPT (Plus) is built upon closed-source large-size LLMs known as the GPT-3.5 and GPT-4. 
\revised{
Much prior research employs medium-size pre-trained models like CodeBERT and CodeT5.
These models are commonly fine-tuned, updating all parameters to align with labeled training data.
~\cite{zhang2020sentiment,zhou2021assessing}.
Though very effective, fine-tuning demands large GPU resources to load and update all parameters of pre-trained models~\cite{Hu2022LoRALA,weyssow2023exploring}. 
As large-size LLMs (e.g., ChatGPT) have a large number of parameters, it is very challenging to fine-tune them using GPU cards widely used in academia.
An alternative and widely adopted approach for large-size LLMs, as introduced in GPT-3, is {\em{in-context learning}} (ICL)~\cite{gpt3,zhou2023patchzero}. 
ICL involves freezing the parameters of LLMs and utilizing suitable prompts to impart task-specific knowledge to the models.  
Unlike fine-tuning, ICL requires no parameter update which significantly reduces the large GPU resource requirement. 
}
To perform the inference/testing, ICL makes predictions based on the probability of generating the next token $t$ given the unlabeled data instance $x$ and the prompt $P$.
Then the output token $t$ is mapped into the prediction categories by the verbalizer (introduced below).

\begin{table}[t]
\caption{Our prompt designs.} 
\vspace{-0.4cm}
\resizebox{0.98\columnwidth}{!}{%
\begin{tabular}{l|l|l|l}
\hline
\textbf{No.} & \textbf{Prompt Type}                                                    & \textbf{Prompt Template}                                                                                                                                                                                            & \multicolumn{1}{c}{\textbf{verbalizer}}                                                \\ \hline
\textbf{P}   & Task Description                                                        & \begin{tabular}[c]{@{}l@{}} Now you need to identify whether a \\ method  contains a vulnerability or not. \\ 
If has any potential vulnerability, \\
output: 'this code is vulnerable'. \\
Otherwise, output: \\
'this code is non-vulnerable'.\\
The code is {[}X{]}. Let's start: {[}Z{]}\end{tabular}                                                                              & \begin{tabular}[c]{@{}l@{}}+: this code is vulnerable\\ -:  this code is non-vulnerable\end{tabular} \\ \hline
\textbf{A1}  & Role Description                                                        & \begin{tabular}[c]{@{}l@{}}You are an experienced developer who \\
knows the security vulnerability very well \end{tabular}                                                                                                                                                                            & \begin{tabular}[c]{@{}l@{}}+: this code is vulnerable\\ -:  this code is non-vulnerable\end{tabular} \\ \hline
\textbf{A2}  & Project Information                                                     & \begin{tabular}[c]{@{}l@{}}The code is from {[}the name of project{]}. \\ The filename is {[}the name of file{]}.\end{tabular}                                                                                      & \begin{tabular}[c]{@{}l@{}}+: this code is vulnerable\\ -:  this code is non-vulnerable\end{tabular} \\ \hline
\textbf{A3}  & \begin{tabular}[c]{@{}l@{}}Dangerous CWE \\ Types Examples\end{tabular} & \begin{tabular}[c]{@{}l@{}}Here are examples of the most dangerous \\ CWE types. \\ Example1: int returnChunkSize ... \\ Lable1: this code is vulnerable. \\ Example2: static ...\\ ...\end{tabular}                       & \begin{tabular}[c]{@{}l@{}}+: this code is vulnerable\\ -:  this code is non-vulnerable\end{tabular} \\ \hline
\textbf{A4}  & \begin{tabular}[c]{@{}l@{}}Randomly Sampled\\ Code\end{tabular}         & \begin{tabular}[c]{@{}l@{}}Here are sampled examples from the \\ training data. \\ Example1: int  ... \\ Lable1: this code is vulnerable. \\ Example2: static ...\\ Lable2: this code is non-vulnerable.\\ ...\end{tabular}       & \begin{tabular}[c]{@{}l@{}}+: this code is vulnerable\\ -:  this code is non-vulnerable\end{tabular} \\ \hline
\textbf{A5}  & \begin{tabular}[c]{@{}l@{}}Retrieved Similar \\ Code\end{tabular}       & \begin{tabular}[c]{@{}l@{}}Here are the most similar codes from \\ the training data. \\ Example1: int  ... \\ Lable1: this code is non-vulnerable. \\ Example2: static ...\\ Lable2: this code is vulnerable.\\ ...\end{tabular} & \begin{tabular}[c]{@{}l@{}}+: this code is vulnerable\\ -:  this code is non-vulnerable\end{tabular} \\ \hline
\end{tabular}
}
\label{tab:prompts}
\vspace{-0.5cm}
\end{table}

\vspace{0.1cm}
\noindent \textbf{Prompt Basics.}
A prompt is a textual string that has two slots: (1) an input slot $[X]$ for the original input data $x$ and (2) an answer slot $[Z]$ for the predicted answer $z$. 
The verbalizer, denoted as $V$, is a function that maps the predicted answer $z$ to a class $\hat{y}$ in the target class set $Y$, formally $V: Z \rightarrow Y$. 
For instance, a straightforward prompt and verbalizer are shown as follows: $ f_{prompt} (x) = `` Code \; is \; [X]. \; It \; is \; [Z]. \; "$ and $ V= \left\{ \begin{array}{lr} +, \; if \; Z=vulnerable, &  \\ -, \; if \; Z=\mathit{non}\text{-}\mathit{vulnerable}, & \end{array} \right.$ where $V$ is the defined verbalizer where the token ``vulnerable'' is mapped into the positive class. 
ChatGPT may generate responses that differ from our predefined label words. To simplify the process, we manually check the predicted classes of ChatGPT's generated answers when they diverge from our specified label words. For example, we map the answer "it is vulnerable because ..." into the "vulnerable" class.

\vspace{0.1cm}
\noindent \textbf{Prompt Designs.}
Table~\ref{tab:prompts} shows our designed prompts (the base prompt $+$ several augmentations). 
\revised{OpenAI allows users to guide ChatGPT through two types of messages/prompts: 1) the system message, influencing ChatGPT’s overall behaviors such as adjusting the personality of ChatGPT, and 2) the user message, containing requests for ChatGPT to address and respond to. We use an empty system message and user messages that are listed as prompts in Table~\ref{tab:prompts}.}
Initially, we designed a straightforward prompt (\textit{P}): \textit{``Now you need to identify whether a method contains a vulnerability or not.''} as the base prompt. This base prompt only briefly describes the task we want LLMs to do.
To provide LLMs with more valuable task-specific information, we propose diverse \textbf{augmentations} (\textit{A*}) to the base prompt, including the following:

\vspace{0.1cm}
\noindent \textit{\underline{Role Description (A1)}}: We explicitly defined the role of LLMs in this task: \textit{``You are an experienced developer who knows the security vulnerability very well''}. This strategy aims to remind LLMs to change their working mode to a security-related one. 

\vspace{0.1cm}
\noindent \textit{\underline{Project Information (A2)}}: Recently, Li et al.~\cite{starcoder} propose the state-of-the-art LLM for code namely StarCoder. Li et al. found that adding the filename in the prompts can substantially improve the effectiveness of StarCoder. We followed them to provide LLMs with the project names and filenames associated with the target code. 

\vspace{0.1cm}
\noindent \textit{\underline{External Source Knowledge (A3)}}: 
The CWE system offers a wealth of information about software vulnerabilities such as code examples of typical vulnerable code.
Leveraging such resources could possibly enhance the prompt generation process for vulnerability detection tasks. In this study, we collected the vulnerable code examples that represent the top 25 most dangerous Common Weakness Enumeration (CWE) types identified in the year 2022~\cite{top25_2022}.
\revised{These examples showcase the characteristics and patterns of vulnerabilities, equipping LLMs with valuable insights. This allows us to extend the model's knowledge beyond the limitations of the training data by leveraging external sources, specifically the CWE system.}

\vspace{0.1cm}
\noindent \textit{\underline{Knowledge in the Training Set (A4)}}: The training data encompasses valuable task-specific knowledge pertinent to a given task. 
However, we can only accommodate a limited number of input-output samples because ChatGPT has a maximum token limit of 4,096. In this strategy, we randomly select K samples from the training data to leverage the knowledge embedded within the training dataset. 
Vulnerable/non-vulnerable samples are both used in this strategy.

\vspace{0.1cm}
\noindent  \textit{\underline{Selective Knowledge in the Training Set (A5)}}: In contrast to the aforementioned strategy, we adopted a different approach by retrieving the top K most similar methods from the training data, instead of randomly sampling. These retrieved methods served as examples to furnish LLMs with pertinent knowledge, aiding their decision-making process when evaluating the test data.
To perform the retrieval process, we employed CodeBERT~\cite{CodeBERT} to transform the code snippets into semantic vectors. Subsequently, we quantified the similarity between two code snippets by calculating the cosine similarity of their respective semantic vectors. 
For a given test code, we retrieved the top K similar methods along with their corresponding vulnerability labels from the training data.

%% file: Sections/03_experiments.tex
\vspace{-0.3cm}
\section{Preliminary Evaluation}

\revised{
In this work, we aim to answer a single research question: 
\textbf{How effective is ChatGPT with different prompt designs in vulnerability detection compared to baselines?}
}

\begin{table*}[t]
\centering
\caption{Results of ChatGPT with diverse prompts and the fine-tuned CodeBERT} 
\vspace{-0.3cm}
\resizebox{1.7\columnwidth}{!}{%
\begin{tabular}{l|c|l|c|r|r|r|r|r}
\hline
\textbf{Model}                                                                         & \textbf{Prompt}                & \textbf{Prompt/Model Description}                                                                                    & \textbf{Param.} & \multicolumn{1}{c|}{\textbf{Accuracy}} & \multicolumn{1}{c|}{\textbf{Precision}} & \multicolumn{1}{c|}{\textbf{Recall}} & \multicolumn{1}{c|}{\textbf{F1 score}} & \multicolumn{1}{c}{\textbf{F0.5 score}} \\ \hline
\multirow{11}{*}{\textbf{\begin{tabular}[c]{@{}l@{}}ChatGPT\\ (GPT-3.5)\end{tabular}}} & \textbf{P}                     & provide the task description to LLM                                                                                  & -               & 50.0                                   & Nan                                     & 0.0                                  & Nan                                    & Nan                                     \\ \cline{2-9} 
                                                                                       & \textbf{P+A1}                  & provide the role description to LLM                                                                                  & -               & 50.0                                   & Nan                                     & 0.0                                  & Nan                                    & Nan                                     \\ \cline{2-9} 
                                                                                       & \textbf{P+A2}                  & provide the project name                                                                                             & -               & 50.0                                   & Nan                                     & 0.0                                  & Nan                                    & Nan                                     \\ \cline{2-9} 
                                                                                       & \textbf{P+A3}                  & \begin{tabular}[c]{@{}l@{}}provide vulnerable code examples from \\ 25 most dangerous CWE Types in 2022\end{tabular} & -               & 59.1                                   & 72.2                                    & 29.5                                 & 41.9                                   & 56.0                                    \\ \cline{2-9} 
                                                                                       & \multirow{3}{*}{\textbf{P+A4}} & \multirow{3}{*}{randomly sample K input-output pairs}                                                                & K=1             & 51.8                                   & 75.2                                    & 6.0                                  & 10.7                                   & 20.5                                    \\ \cline{4-9} 
                                                                                       &                                &                                                                                                                      & K=3             & 58.8                                   & 65.8                                    & 41.2                                 & 50.2                                   & 58.4                                    \\ \cline{4-9} 
                                                                                       &                                &                                                                                                                      & K=5             & 61.4                                   & \textbf{80.0}                           & 30.3                                 & 43.9                                   & 60.1                                    \\ \cline{2-9} 
                                                                                       & \multirow{3}{*}{\textbf{P+A5}} & \multirow{3}{*}{retrieve top K most similar code}                                                                    & K=1             & 55.4                                   & 67.2                                    & 21.2                                 & 32.3                                   & 46.9                                    \\ \cline{4-9} 
                                                                                       &                                &                                                                                                                      & K=3             & 56.7                                   & 60.1                                    & 39.9                                 & 48.0                                   & 54.6                                    \\ \cline{4-9} 
                                                                                       &                                &                                                                                                                      & K=5             & 59.8                                   & 63.2                                    & 47.2                                 & 54.0                                   & 59.2                                    \\ \cline{2-9} 
                                                                                       & \textbf{P+A4+A5}               & \begin{tabular}[c]{@{}l@{}}combine P4 and P5 together (both top 3)\end{tabular}                            & K=3             & \textbf{62.7}                          & 76.3                                    & 36.8                                 & 49.7                                   & \textbf{62.8}                           \\ \hline  
\textbf{CodeBERT}                                                                      & \textbf{-}                     & full-parameter fine-tuned                                                                                          & -               & 60.3                                   & 62.3                                    & \textbf{53.3}                        & \textbf{57.3}                          & 60.1                                   \\ \hline 
\end{tabular}
}
\label{tab:results}
\vspace{-0.1cm}
\end{table*}

\begin{table*}[h]
\centering
\caption{Results of GPT-4 and baselines on the first half of the test set} 
\vspace{-0.3cm}
\resizebox{1.7\columnwidth}{!}{%
\begin{tabular}{l|c|l|c|r|r|r|r|r}
\hline
\textbf{Model}                                                                         & \textbf{Prompt}                & \textbf{Prompt/Model Description}                                                                                    & \textbf{Param.} & \multicolumn{1}{c|}{\textbf{Accuracy}} & \multicolumn{1}{c|}{\textbf{Precision}} & \multicolumn{1}{c|}{\textbf{Recall}} & \multicolumn{1}{c|}{\textbf{F1 score}} & \multicolumn{1}{c}{\textbf{F0.5 score}} \\ \hline
\textbf{ \textbf{\begin{tabular}[c]{@{}l@{}} ChatGPT\\ (GPT-3.5) \end{tabular}}}                                                                      & \textbf{P+A4+A5}                     & combine P4 and P5 together                                                                                          & K=3               & 63.6                                 & \textbf{77.8}                                  & 38.0                         & 51.1                          &  64.3                                 \\ \hline
\multirow{2}{*}{\textbf{\begin{tabular}[c]{@{}l@{}}\\ ChatGPT\\ (GPT-4) \end{tabular}}} & \textbf{P}                     & provide the task description to LLM                                                                                  & -               & 60.3                                   & 67.3                                     & 40.2                                  & 50.3                                    & 59.3                                    \\ \cline{2-9} 
& \textbf{P+A3}   & code examples from CWE Types                                                                                & -               & \textbf{75.5}                                   &  73.7                                    & \textbf{79.3}                                  &  \textbf{76.4}                                 &  \textbf{74.8}                                   \\ \cline{2-9}
& \textbf{P+A5}   & retrieve top K most similar code                                                                                  & K=5               &61.4                                    &63.6                                      &53.3                                  &58.0                                 &61.3                                    \\ \cline{2-9} 
& \textbf{P+A4+A5}   & combine A4 and A5 together                                                                             & K=3               & 59.2                                   &  60.2                                    &  54.3                                 & 57.1                                 & 59.0                                   \\  \hline
\textbf{CodeBERT}                                                                      & \textbf{-}                     & fine-tuned, tested on a half test set                                                                                          & -               & 56.0                                   & 57.3                                    & 46.7                        & 51.5                          & 54.8                                   \\ \hline
\end{tabular}
}
\label{tab:results2}
\vspace{-0.3cm}
\end{table*}

\vspace{0.08cm}
\noindent \textbf{Dataset and Model.}
We use the vulnerability-fixing commit dataset recently collected by Pan et al.~\cite{treevul}.
To get the vulnerable functions from vulnerability-fixing commits, we followed Fan et al.~\cite{fan2020ac} to first collect software versions prior to a vulnerability-fixing commit, and then labeled functions with lines changed in a patch as vulnerable. All remaining functions in a file touched by a commit were regarded as non-vulnerable.
As the security patches dataset covers a large number of software repositories implemented in diverse programming languages, it is challenging to write all the corresponding parsers (used to split functions from files) for all languages. Thus, in this preliminary evaluation, we only focus on software repositories implemented in \textit{C/C++}. 
To build our test set, we first randomly sampled 20 open-source software repositories implemented in C/C++ from the original test set split by Pan et al.~\cite{treevul} and used their vulnerability fixes to get the vulnerable functions (positive samples) for our test set. 
Due to the considerable cost~\cite{openai_price}
associated with querying ChatGPT on a large test set, we limited our sampling to 20 repositories. Despite the restricted sample size, our preliminary experiment was designed to showcase the potential of ChatGPT. 
For our training/validation sets, we used all the vulnerability fixes of the C/C++ repositories in the original training/validation sets split by Pan et al. and extracted vulnerable functions (positive samples). 
To obtain negative samples, i.e., non-vulnerable functions, for the test/training/validation sets, we employed a random sampling technique. For each vulnerable function, we selected one function at random from non-vulnerable functions that were extracted from the same file as the vulnerable function. 
Different from the vulnerable function, those non-vulnerable functions had not been modified by the vulnerability-fixing commit. 
Finally, our dataset has 7,683/853/368 methods in the training/validation/test sets.

For studied models, we primarily focused our investigation on ChatGPT (GPT-3.5) with the model name {\tt gpt-3.5-turbo} while also doing limited experiments on ChatGPT (GPT-4). 
For the baseline model, we opted for one of the state-of-the-art approaches (i.e., CodeBERT) according to a recent comprehensive empirical study~\cite{steenhoek2022empirical}.

\vspace{0.1cm}
\noindent \textbf{Evaluation.}
To measure the model's effectiveness, we adopted widely used evaluation metrics,  i.e., Accuracy, Precision, Recall, F1, and F0.5. We incorporated the F0.5 metric, which assigns greater importance to precision than to recall. 
This choice is motivated by the developers' aversion to false positives, as a low success rate may diminish their patience and confidence in the system~\cite{kochhar2016practitioners}. 
As the ICL method exhibits some instability, in this preliminary work, we repeated experiments twice and reported the average results.

\vspace{0.1cm}
\noindent \textbf{Results.} 
The performance of GPT-3.5 in vulnerability detection was evaluated by integrating different prompts, and the results are summarized in Table~\ref{tab:results}. 
Experimental results revealed that the base prompt yielded unsatisfactory outcomes, with GPT-3.5 predicting every target code as non-vulnerable, resulting in an accuracy of 50\% and a recall of 0\%.
The inclusion of role descriptions and project information did not contribute to better performance. However, incorporating examples from external source knowledge, specifically the 25 most dangerous CWE types, led to substantial performance improvements (18.2\% in Accuracy).
Furthermore, the utilization of random sampling codes from the training data and retrieving similar codes also resulted in significantly better performance (up to 22.8\% and 19.6\% in Accuracy) compared to the base prompt. Among all the prompt combinations studied, the {\tt P+A5} combination achieved the highest F1 score (54.0\%) and Recall (47.2\%), and the {\tt P+A4+A5} combination achieved the best F0.5 (62.8\%) score and Accuracy (62.7\%). 
The {\tt P+A4+A5} combination outperforms the base prompt {\tt P} by 25.4\% in Accuracy.
When comparing GPT-3.5 to the state-of-the-art approach, GPT-3.5 ({\tt P+A4+A5}) outperformed CodeBERT by 4.0\%, 22.5\%, and 4.5\% in terms of Accuracy, Precision, and F0.5, respectively. However, GPT-3.5 underperformed CodeBERT by 44.8\% and 15.3\% in Recall and F1. 
These experimental results highlight the distinct strengths of CodeBERT and GPT-3.5. 
GPT-3.5 demonstrates significantly higher Precision scores, indicating its proficiency in minimizing false positives. On the other hand, CodeBERT showcases a much higher Recall score, signifying its capability to identify a greater number of vulnerabilities. \textbf{Overall, GPT-3.5 demonstrates competitive performance when compared to the fine-tuned CodeBERT.}

The performance of GPT-4 is presented in Table~\ref{tab:results2}. 
Notably, GPT-4 is accompanied by a considerably higher cost.
Due to the high costs, in this preliminary evaluation, we only evaluate GPT-4 in the first half of the test set. To assess its performance, we employed four different prompts: the base prompt ({\tt P}), the external source knowledge prompt {\tt P+A3}, the prompt {\tt P+A5} which obtained the second best F1 for GPT-3.5, and the prompt {\tt P+A4+A5} which obtained the best F1 for GPT-3.5.
As illustrated in Table~\ref{tab:results2}, 
\textbf{GPT-4 with the prompt {\tt \textbf{P+A3}} significantly outperformed the fine-tuned CodeBERT by 34.8\% in terms of accuracy}.

%% file: Sections/04_threats.tex
\vspace{-0.2cm}
\section{Threats to validity}

One concern arises due to the possibility of data leakage in ChatGPT. 
The dataset used in our evaluation may overlap with the data used for training ChatGPT. 
However, because ChatGPT is a closed-source model, we lack the means to validate whether such an overlap exists.
\revised{Another threat to validity is the equal data ratio between vulnerable and non-vulnerable functions in the test set. We create a balanced test set to alleviate costs linked to ChatGPT usage, given that the expenses of these experiments rise in proportion to the number of test samples, particularly with a large number of non-vulnerable functions. However, the equal data ratio does not realistically represent the actual problem of vulnerability prediction where the vulnerable code is the minority in software systems. Besides, this data ratio will lead to inflated metrics in this study. We recognize this issue as a limitation of the preliminary study, and in our future work, we aim to develop a new test set that accurately reflects the real-world data ratio between vulnerable and non-vulnerable functions.}
\revised{Lastly, we followed Fan et al.~\cite{fan2020ac} to label functions with lines changed in a vulnerability-fixing commit as vulnerable. This labeling heuristic may overestimate the actual number of vulnerable functions due to the commit tangling, where some changed functions in the vulnerability-fixing commit may not be directly relevant to addressing the vulnerability. 
We recognize the challenge of accurately identifying real vulnerability-related functions in the vulnerability-fixing commit. In our future work, we aim to employ existing techniques (e.g.,~\cite{li2022utango}) to untangle the vulnerability-fixing commits.
}

%% file: Sections/related_work.tex
\vspace{-0.2cm}
\section{Related Work}

\revisedtwo{
We are unable to find vulnerability detection work using GPT-3.5 and GPT-4 in the academic literature. However, there are several related studies in the gray (non-peer-reviewed) literature. For example, BurpGPT~\cite{burpGPT} integrates ChatGPT with the Burp Suite~\cite{burp} to detect vulnerabilities in web applications. In contrast, the software studied in our study is not confined to a specific domain. Also, vuln\_GPT, an LLM designed to discover and address software vulnerabilities, was introduced recently~\cite{vul_GPT}. Different from~\cite{vul_GPT}, our study focuses on improving prompts for vulnerability detection. A parallel work~\cite{zhang2023prompt} also explores the use of ChatGPT in vulnerability detection. They mainly enhance prompts through structural and sequential auxiliary information. Three distinctions of our work from~\cite{zhang2023prompt} are: 1) We also studied GPT-4 while they did not; 2) We integrate knowledge from the CWE system and similar samples from the training set to enhance prompts, a dimension not considered in~\cite{zhang2023prompt}; 3) We identified promising future study directions which are not thoroughly discussed in~\cite{zhang2023prompt}.
}

%% file: Sections/05_future_work.tex
\vspace{-0.2cm}
\section{Future Work}

\revised{
There are plenty of exciting paths to explore in future research. Here, we're just highlighting a few of these potential directions:}

\vspace{0.08cm}
\noindent
\revised{
\textbf{Local and Specialized LLMs-based Vulnerability Detection.} 
This study focuses on ChatGPT. However, ChatGPT requires data to be sent to third-party services. This may restrict the utilization of ChatGPT-related vulnerability detection tools among specific organizations, such as major tech corporations or governments. These organizations, e.g.,~\cite{us_dos_report1, report2, report3}, regard their source code as proprietary, sensitive, or classified material, and as a result, they are unable to transmit or share it with third-party services. Additionally, ChatGPT models are not specialized for vulnerability detection and thus may not take full advantage of the rich open-source vulnerability data.
}

\revised{
To address the above-mentioned limitations, in the future, we aim to propose a local and specialized LLM solution for vulnerability detection.
The solution will build upon general-purpose and open-source code LLMs, e.g., Llama~\cite{code_llama}, which will be tuned for vulnerability detection with the relevant vulnerability corpus. The tuned LLM can alleviate the concerns of organizations that prioritize data security and privacy while also making use of the abundant open-source vulnerability data.
}

\vspace{0.08cm}
\noindent
\revised{
\textbf{Precision and Robustness Boost in Vulnerability Detection.} A vulnerability detection solution with a high precision is usually preferred. Additionally, the solution needs to remain robust against data perturbations or adversarial attacks. 
With higher precision, developers will have greater confidence in the reliability of detections and consequently perform the requisite actions to address the detected vulnerabilities. 
With higher robustness, a more secure and stable vulnerability detection model can be produced, and be more immune to adversarial attacks. }

\revised{
In the future, we plan to boost LLMs' precision and robustness in this task: 1) To improve precision, we plan to employ ensemble learning, a promising technique to improve the precision (and possibly recall) by identifying the common high-confidence predictions among different models. 2) To improve robustness, we plan to extend an existing work~\cite{yang2022natural} that employs a single adversarial transformation (renaming of variables) to enhance model robustness. We will delve into various other types of adversarial transformations and assess their effectiveness in enhancing the robustness of LLMs.}

\vspace{0.08cm}
\noindent
\revised{
\textbf{Enhancing Effectiveness in Long-Tailed Distribution.} In this study, we formulate the task as a binary classification (vulnerable or non-vulnerable). Moving one step further, developers may require the tool to indicate the specific vulnerability type (e.g., CWE types) associated with the detected vulnerable code. This additional information is crucial for a better understanding and resolution of the vulnerability.
However, a recent study~\cite{zhou2023devil} revealed that vulnerability data exhibit a long-tailed distribution in terms of CWE types: a small number of CWE types have a substantial number of samples, while numerous CWE types have very few samples. The study also pointed out that LLMs struggle to effectively handle vulnerabilities in these less common types. This long-tailed distribution could pose a challenge for LLMs-based vulnerability detection solutions.
}

\revised{
In the future, we plan to 1) explore whether LLMs, specifically ChatGPT, can effectively detect these infrequent vulnerabilities or not and 2) propose a solution (e.g., generating more samples for the less common types via data augmentation) to address the impact of the long-tailed distribution of vulnerability data. 
}

\vspace{0.08cm}
\noindent
\revisedzx{
\textbf{Trust and Synergy with Developers.} AI-powered solutions for vulnerability detection, including this work, have limited interaction with developers. They may face challenges in establishing trust and synergy with developers during practical use. To overcome this, future works should investigate more effective strategies to foster trust and collaboration between developers and AI-powered solutions~\cite{DBLP:journals/corr/abs-2309-04142}. By nurturing trust and synergy, AI-powered solutions may evolve into smart workmates to better assist developers.}

%% file: Sections/06_conclusions.tex
\vspace{-0.2cm}
\section{Conclusion}
\label{sec:conclusion}
In this study, we explored the efficacy and potential of LLMs (i.e., ChatGPT) in vulnerability detection.
We proposed some insightful prompt enhancements such as incorporating the external knowledge and choosing valuable samples from the training set. 
We also identified many promising directions for future study.
We made our replication package\footnote{\url{https://github.com/soarsmu/ChatGPT-VulDetection}} publicly available for future studies.

\vspace{0.2cm}
\noindent \textbf{Acknowledgement.} \revised{ This research / project is supported by the National Research Foundation, under its Investigatorship Grant (NRF-NRFI08-2022-0002). Any opinions, findings and conclusions or recommendations expressed in this material are those of the author(s) and do not reflect the views of National Research Foundation, Singapore.}

\balance

%% file: main.bbl
\begin{thebibliography}{10}

\bibitem{MSFlaw}
{Microsoft Exchange Flaw: Attacks Surge After Code Published.}
\newblock \url{https://www.bankinfosecurity.com/ms-exchange-flaw-causes-spike-intrdownloader-gen-trojans-a-16236}, 2022.

\bibitem{turner2008symantec}
Dean Turner, Marc Fossi, Eric Johnson, Trevor Mack, Joseph Blackbird, Stephen Entwisle, Mo~King Low, David McKinney, and Candid Wueest.
\newblock Symantec global internet security threat report--trends for july-december 07.
\newblock {\em Symantec Enterprise Security}, 13:1--36, 2008.

\bibitem{hanif2022vulberta}
Hazim Hanif and Sergio Maffeis.
\newblock Vulberta: Simplified source code pre-training for vulnerability detection.
\newblock In {\em 2022 International Joint Conference on Neural Networks (IJCNN)}, pages 1--8. IEEE, 2022.

\bibitem{fu2022linevul}
Michael Fu and Chakkrit Tantithamthavorn.
\newblock Linevul: a transformer-based line-level vulnerability prediction.
\newblock In {\em Proceedings of the 19th International Conference on Mining Software Repositories}, pages 608--620, 2022.

\bibitem{nguyen2022regvd}
Van-Anh Nguyen, Dai~Quoc Nguyen, Van Nguyen, Trung Le, Quan~Hung Tran, and Dinh Phung.
\newblock Regvd: Revisiting graph neural networks for vulnerability detection.
\newblock In {\em Proceedings of the ACM/IEEE 44th International Conference on Software Engineering: Companion Proceedings}, pages 178--182, 2022.

\bibitem{zhou2019devign}
Yaqin Zhou, Shangqing Liu, Jingkai Siow, Xiaoning Du, and Yang Liu.
\newblock Devign: Effective vulnerability identification by learning comprehensive program semantics via graph neural networks.
\newblock {\em Advances in neural information processing systems}, 32, 2019.

\bibitem{CodeBERT}
Zhangyin Feng, Daya Guo, Duyu Tang, Nan Duan, Xiaocheng Feng, Ming Gong, Linjun Shou, Bing Qin, Ting Liu, Daxin Jiang, and Ming Zhou.
\newblock Codebert: {A} pre-trained model for programming and natural languages.
\newblock In Trevor Cohn, Yulan He, and Yang Liu, editors, {\em Findings of the Association for Computational Linguistics: {EMNLP} 2020, Online Event, 16-20 November 2020}, volume {EMNLP} 2020 of {\em Findings of {ACL}}, pages 1536--1547. Association for Computational Linguistics, 2020.

\bibitem{xia2023keep}
Chunqiu~Steven Xia and Lingming Zhang.
\newblock Keep the conversation going: Fixing 162 out of 337 bugs for \$0.42 each using chatgpt.
\newblock In {\em arXiv preprint arXiv:2304.00385}, 2023.

\bibitem{zhang2023cupid}
Ting Zhang, Ivana~Clairine Irsan, Ferdian Thung, and David Lo.
\newblock Cupid: Leveraging chatgpt for more accurate duplicate bug report detection.
\newblock {\em arXiv preprint arXiv:2308.10022}, 2023.

\bibitem{zhang2023revisiting}
Ting Zhang, Ivana~Clairine Irsan, Ferdian Thung, and David Lo.
\newblock Revisiting sentiment analysis for software engineering in the era of large language models.
\newblock {\em arXiv preprint arXiv:2310.11113}, 2023.

\bibitem{weyssow2023exploring}
Martin Weyssow, Xin Zhou, Kisub Kim, David Lo, and Houari Sahraoui.
\newblock Exploring parameter-efficient fine-tuning techniques for code generation with large language models.
\newblock {\em arXiv preprint arXiv:2308.10462}, 2023.

\bibitem{zhou2023patchzero}
Xin Zhou, Bowen Xu, Kisub Kim, DongGyun Han, Thanh Le-Cong, Junda He, Bach Le, and David Lo.
\newblock Patchzero: Zero-shot automatic patch correctness assessment.
\newblock {\em arXiv preprint arXiv:2303.00202}, 2023.

\bibitem{zhang2020sentiment}
Ting Zhang, Bowen Xu, Ferdian Thung, Stefanus~Agus Haryono, David Lo, and Lingxiao Jiang.
\newblock Sentiment analysis for software engineering: How far can pre-trained transformer models go?
\newblock In {\em 2020 IEEE International Conference on Software Maintenance and Evolution (ICSME)}, pages 70--80. IEEE, 2020.

\bibitem{zhou2021assessing}
Xin Zhou, DongGyun Han, and David Lo.
\newblock Assessing generalizability of codebert.
\newblock In {\em 2021 IEEE International Conference on Software Maintenance and Evolution (ICSME)}, pages 425--436. IEEE, 2021.

\bibitem{Hu2022LoRALA}
Edward~J. Hu, Yelong Shen, Phillip Wallis, Zeyuan Allen-Zhu, Yuanzhi Li, Shean Wang, and Weizhu Chen.
\newblock Lora: Low-rank adaptation of large language models.
\newblock {\em ArXiv}, abs/2106.09685, 2022.

\bibitem{gpt3}
Tom Brown, Benjamin Mann, Nick Ryder, Melanie Subbiah, Jared~D Kaplan, Prafulla Dhariwal, Arvind Neelakantan, Pranav Shyam, Girish Sastry, Amanda Askell, et~al.
\newblock Language models are few-shot learners.
\newblock {\em Advances in neural information processing systems}, 33:1877--1901, 2020.

\bibitem{starcoder}
Raymond Li, Loubna~Ben Allal, Yangtian Zi, Niklas Muennighoff, Denis Kocetkov, Chenghao Mou, Marc Marone, Christopher Akiki, Jia Li, Jenny Chim, et~al.
\newblock Starcoder: may the source be with you!
\newblock {\em arXiv preprint arXiv:2305.06161}, 2023.

\bibitem{top25_2022}
\url{https://cwe.mitre.org/top25/archive/2022/2022_cwe_top25.html}, 2022.

\bibitem{treevul}
Shengyi Pan, Lingfeng Bao, Xin Xia, David Lo, and Shanping Li.
\newblock Fine-grained commit-level vulnerability type prediction by cwe tree structure.
\newblock In {\em 45th International Conference on Software Engineering, {ICSE} 2023}, 2023.

\bibitem{fan2020ac}
Jiahao Fan, Yi~Li, Shaohua Wang, and Tien~N Nguyen.
\newblock Ac/c++ code vulnerability dataset with code changes and cve summaries.
\newblock In {\em Proceedings of the 17th International Conference on Mining Software Repositories}, pages 508--512, 2020.

\bibitem{openai_price}
\url{https://openai.com/pricing}, 2023.

\bibitem{steenhoek2022empirical}
Benjamin Steenhoek, Md~Mahbubur Rahman, Richard Jiles, and Wei Le.
\newblock An empirical study of deep learning models for vulnerability detection.
\newblock {\em 45th International Conference on Software Engineering, {ICSE}}, 2023.

\bibitem{kochhar2016practitioners}
Pavneet~Singh Kochhar, Xin Xia, David Lo, and Shanping Li.
\newblock Practitioners' expectations on automated fault localization.
\newblock In {\em Proceedings of the 25th International Symposium on Software Testing and Analysis}, pages 165--176, 2016.

\bibitem{li2022utango}
Yi~Li, Shaohua Wang, and Tien~N Nguyen.
\newblock Utango: untangling commits with context-aware, graph-based, code change clustering learning model.
\newblock In {\em Proceedings of the 30th ACM Joint European Software Engineering Conference and Symposium on the Foundations of Software Engineering}, pages 221--232, 2022.

\bibitem{burpGPT}
Alexandre Teyar.
\newblock Burpgpt - chatgpt powered automated vulnerability detection tool.
\newblock \url{https://burpgpt.app/#faq}, 2023.

\bibitem{burp}
PortSwigger.
\newblock Burp suite - application security testing software.
\newblock \url{https://portswigger.net/burp}.

\bibitem{vul_GPT}
Vicarius.
\newblock vuln\_gpt debuts as ai-powered approach to find and remediate software vulnerabilities.
\newblock \url{https://venturebeat.com/ai/got-vulns-vuln_gpt-debuts-as-ai-powered-approach-to-find-and-remediate-software-vulnerabilities/}, 2023.

\bibitem{zhang2023prompt}
Chenyuan Zhang, Hao Liu, Jiutian Zeng, Kejing Yang, Yuhong Li, and Hui Li.
\newblock Prompt-enhanced software vulnerability detection using chatgpt.
\newblock {\em arXiv preprint arXiv:2308.12697}, 2023.

\bibitem{us_dos_report1}
Jon Harper.
\newblock {Pentagon testing generative AI in ‘global information dominance’ experiments.}
\newblock \url{https://defensescoop.com/2023/07/14/pentagon-testing-generative-ai-in-global-information-dominance-experiments/}, 2023.

\bibitem{report2}
Kyle Chua.
\newblock {Samsung Bans Use of Generative AI Tools on Company-Owned Devices Over Security Concerns.}
\newblock \url{https://www.tech360.tv/samsung-bans-use-generative-ai-tools}, 2023.

\bibitem{report3}
Kyle Chua.
\newblock {Apple Bans Internal Use of ChatGPT and GitHub Copilot Over Fear of Leaks}.
\newblock \url{https://www.tech360.tv/apple-bans-internal-use-chatgpt-github-copilot-over-fears-of-leaks}, 2023.

\bibitem{code_llama}
Hugo Touvron, Louis Martin, Kevin Stone, Peter Albert, Amjad Almahairi, Yasmine Babaei, Nikolay Bashlykov, Soumya Batra, Prajjwal Bhargava, Shruti Bhosale, et~al.
\newblock Llama 2: Open foundation and fine-tuned chat models.
\newblock {\em arXiv preprint arXiv:2307.09288}, 2023.

\bibitem{yang2022natural}
Zhou Yang, Jieke Shi, Junda He, and David Lo.
\newblock Natural attack for pre-trained models of code.
\newblock In {\em Proceedings of the 44th International Conference on Software Engineering}, 2022.

\bibitem{zhou2023devil}
Xin Zhou, Kisub Kim, Bowen Xu, Jiakun Liu, DongGyun Han, and David Lo.
\newblock The devil is in the tails: How long-tailed code distributions impact large language models.
\newblock {\em arXiv preprint arXiv:2309.03567}, 2023.

\bibitem{DBLP:journals/corr/abs-2309-04142}
David Lo.
\newblock Trustworthy and synergistic artificial intelligence for software engineering: Vision and roadmaps.
\newblock {\em CoRR}, abs/2309.04142, 2023.

\end{thebibliography}
